\documentclass[a4paper,12pt]{article}

\usepackage{amsthm}
\usepackage{amsmath,amssymb,latexsym,amsfonts,mathrsfs}
\usepackage[dvipdfmx]{graphicx}

\usepackage{color}

\usepackage{multicol}

\usepackage{fancyhdr}
\usepackage[top=30truemm, bottom=30truemm, left=25truemm, right=25truemm]{geometry}

\newcommand{\ep}{\varepsilon}
\newcommand{\nn}{\nonumber}

\newcommand{\SCR}[1]{{\mathscr #1}}

\newcommand{\CAL}[1]{{\cal #1}
}

\newcommand{\MAT}[1]{\left(\begin{array}{cccccccccc}#1\end{array}\right)}

\newcommand{\J}[1]{\left\langle #1 \right\rangle}
\newcommand{\D}[1]{{\mathscr D}( #1 )}

\newcommand{\bfR}{ {\bf R} }
\newcommand{\bfN}{ {\bf N} }

\theoremstyle{definition}
\newtheorem{Thm}{{\bf Theorem}}[section]

\newtheorem{Lem}[Thm]{{\bf Lemma}}
\newtheorem{Prop}[Thm]{{\bf Proposition}}
\newtheorem{Cor}[Thm]{{\bf Corollary}}
\newtheorem{Ass}[Thm]{{\bf Assumption}}

\newtheorem{Rem}[Thm]{{\bf Remark}}

\newcounter{Exami}

\newcommand{\Proof}[2][Proof]{
\begin{proof}[{\bf #1}]
#2
\end{proof}
}

\begin{document}

%%%%%%%%%%%%%%%%%%%%%%%%%%%%%%%%%%%%%%%%%%%%%%%%%%%%%%%%%%%%%%%%%%%%%%%%%%%%%%%%%%%%%%%%%%%%%%%%%%%%%%%%%%%%%%%%%%%%%%%%%%%
%author
%%%%%%%%%%%%%%%%%%%%%%%%%%%%%%%%%%%%%%%%%%%%%%%%%%%%%%%%%%%%%%%%%%%%%%%%%%%%%%%%%%%%%%%%%%%%%%%%%%%%%%%%%%%%%%%%%%%%%%%%%%%
\begin{flushleft}
{\bf \Large Asymptotic completeness of wave operators for Schr\"{o}dinger operators with time-periodic magnetic fields} \\ \vspace{0.3 cm} 
by 
{\bf \large  Masaki Kawamoto.} \\  
Graduate School of Science and Engineering, Ehime University, 3 Bunkyo-cho Matsuyama, Ehime 790-8577. Japan. \\ 
Email: kawamoto.masaki.zs@ehime-u.ac.jp, 
\end{flushleft}

%%%%%%%%%%%%%%%%%%%%%%%%%%%%%%%%%%%%%%%%%%%%%%%%%%%%%%%%%%%%%%%%%%%%%%%%%%%%%%%%%%%%%%%%%%%%%%%%%%%%%%%%%%%%%%%%%%%%%%%%%%%
%abst
%%%%%%%%%%%%%%%%%%%%%%%%%%%%%%%%%%%%%%%%%%%%%%%%%%%%%%%%%%%%%%%%%%%%%%%%%%%%%%%%%%%%%%%%%%%%%%%%%%%%%%%%%%%%%%%%%%%%%%%%%%%
\begin{center}
\begin{minipage}[c]{400pt}
{\bf Abstract} {\small Under the effect of suitable time-periodic magnetic fields, the velocity of a charged particle grows exponentially in $t$; this phenomenon provides the asymptotic completeness for wave operators with slowly decaying potentials. These facts were shown under some restrictions for time-periodic magnetic fields and the range of wave operators. In this study, we relax these restrictions and finally obtain the asymptotic completeness of wave operators. Additionally, we show them under generalized conditions, which are truly optimal for time-periodic magnetic fields. Moreover, we provide a uniform resolvent estimate for the perturbed Floquet Hamiltonian.  
}
\end{minipage}
\end{center}

\begin{flushleft}
\end{flushleft}
\begin{flushleft}
{\bf MSC classification 2020}: Primary 81U05; Second 35P25, 47A55. 
\end{flushleft}
%%%%%%%%%%%%%%%%%%%%%%%%%%%%%%%%%%%%%%%%%%%%%%%%%%%%%%%%%%%%%%%%%%%%%%%%%%%%%%%%%%%%%%%%%%%%%%%%%%%%%%%%%%%%%%%%%%%%%%%%%%%
%Intro
%%%%%%%%%%%%%%%%%%%%%%%%%%%%%%%%%%%%%%%%%%%%%%%%%%%%%%%%%%%%%%%%%%%%%%%%%%%%%%%%%%%%%%%%%%%%%%%%%%%%%%%%%%%%%%%%%%%%%%%%%%%

\section{Introduction}

We study a scattering problem for a charged particle under the effect
of a time-periodic magnetic field. In this study, we assume that the
charged particle moves on a plane ${\bf R} ^2$ in the presence of a time-periodic magnetic field ${\bf B}
(t) = (0,0,B(t))$ with $B(t+T) = B(t)$, which is always perpendicular to the plane. Subsequently, the free Hamiltonian
for this system is given by
\begin{align*} %\label{2}
H_0(t) = (p-qA(t,x))^2/(2m), \quad 
A(t,x) = (-B(t)x_2 , B(t)x_1)/2,
\end{align*}
where $x=(x_1,x_2) \in {\bf R} ^2$, 
$p=(p_1, p_2) = -i (\partial_1 , \partial _2)$, $m>0$ and $q \in \bfR \backslash
\{0\}$ are position, momentum, mass, and charge of a particle, respectively. 
$B(t) \in L^{\infty}({\bf R})$ is the intensity of the magnetic field at time $t$. 
The wave function $\psi(t,x)$ described by this system 
satisfies the following time-dependent Schr\"{o}dinger equation; 
\begin{align*}%\label{1}
\begin{cases}
i \partial _t \psi(t,x) &= H_0(t) \psi(t,x), \\
\psi(0,x) &= \psi_0
\end{cases}, 
\end{align*}
By defining a propagator for $H_0(t)$ as $U_0(t,s)$, the wave function $\psi(t,x)$ is
denoted by $\psi(t,x) = U_0(t,0) \psi_0$, which we refer to as family of unitary operators $\{U_0(t,s)\}_{(t,s) \in {\bf R}^2}$ a propagator for $H_0(t)$ 
if each component satisfies 
\begin{align*}
&i \partial _t U_0(t,s) =H_0(t)U_0(t,s) , \quad 
i \partial_s U_0(t,s) = - U_0(t,s) H_0(s) \\ 
&U_0(t,\theta)U_0(\theta ,s) =U_0(t,s), \quad U_0(s,s)= \mathrm{Id}_{L^2({\bf R}^2)}.
\end{align*} 
Under these settings, the classical trajectory $x(t)$ and $p(t)$ of this system can be expressed in the form 
\begin{align*}
x(t) := U_0(0,t) x U_0(t,0), \quad p(t) := U_0(0,t) p U_0(t,0).
\end{align*}
Let us define $L := x_1p_2-x_2p_1$, $\omega (t) = qB(t)/m$, and $\Omega (t) = {\displaystyle \int_0^t \omega (s)} ds$. 
It can be noted that $\tilde{U}_0(t,0):= e^{-i\Omega (t) L} U_0(t,0)$ is a propagator for $\tilde{H}_0(t) := p^2/(2m) + q^ 2 B ^2(t)x^2/(8m)$ because 
$L$ commutes with $p^2 $ and $x^2$. By defining $\tilde{x}(t)$ and $\tilde{p}(t)$ as $\tilde{U}_0(0,t) x \tilde{U}_0(t,0)$ and 
$\tilde{U}_0(0,t) p \tilde{U}_0 (t,0)$, respectively, the straightforward calculation shows that
\begin{align*}
& \tilde{x}'(t) = \tilde{U}_0(0,t) i[\tilde{H}_0(t), x] \tilde{U}_0(t,0) = \tilde{p}(t)/m, \\ 
& \tilde{p}'(t) /m = \tilde{U}_0(0,t) i[\tilde{H}_0(t), p/m] \tilde{U}_0(t,0) = -(qB(t)/(2m))^2 \tilde{x}(t).
\end{align*}
hold on $\SCR{S}({\bf R}^2)$, where $[\cdot, \cdot ] $ denotes the commutator of operators, and these equations yield {\em Hill's equation }
\begin{align*}
\tilde{x}''(t) + \left( \frac{qB(t)}{2m} \right) ^2 \tilde{x}(t) =0 , \quad 
\begin{cases}
\tilde{x}(0) = 0, \\ 
\tilde{x}'(0) = \tilde{p}(0)/m.
\end{cases}
\end{align*}
and a differential equation 
\begin{align*}
\tilde{p}(t)=m\tilde{x} '(t), 
\end{align*}
refer to Kawamoto \cite{Ka} \S{3}. Hence, by introducing the {\em fundamental solutions of the Hill's equation}, $\zeta _1 (t)$ and $\zeta _2 (t)$ as 
\begin{align}\label{14} 
\zeta _j ''(t) + \left( \frac{qB(t)}{2m} \right) ^2 \zeta _j(t) =0, \quad 
\begin{cases}
\zeta _1(0) = 1, \\ 
\zeta _1 '(0) = 0, 
\end{cases}
\quad 
\begin{cases}
\zeta _2 (0) = 0, \\ 
\zeta _2 '(0) = 1,
\end{cases}
\end{align}
we obtain 
\begin{align*}
\MAT{\tilde{x}(t) \\ \tilde{p}(t) } = 
\MAT{\zeta _1 (t) & \zeta _2 (t)/m \\ m \zeta _1 '(t) & \zeta _2' (t)} \MAT{x \\ p}.
\end{align*}
Noting 
\begin{align*}
e^{-i\Omega (t) L} \MAT{x \\ p} e^{i\Omega (t) L} = \MAT{\hat{R}(\Omega (t)/2) x \\ \hat{R}(\Omega (t)/2) p}, \quad. 
\hat{R}(t) = \MAT{\cos t & \sin t \\ - \sin t & \cos t},  
\end{align*}
(refer to Adachi-Kawamoto \cite{AK} or \cite{Ka}), it can be deduced that 
\begin{align*}
\MAT{x(t) \\ p(t)} = \MAT{\zeta _1 (t) & \zeta _2 (t)/m \\ m \zeta _1 '(t) & \zeta _2' (t)} \MAT{\hat{R}(\Omega (t)/2)x \\ \hat{R}(\Omega (t)/2)p} 
\end{align*}
holds. Here we let 
\begin{align*}
\CAL{L} := \MAT{ \zeta _1 (T) & \zeta _2 (T) /m \\ m \zeta _1 '(T) & \zeta _2' (T) }. 
\end{align*}
Then, for $t = NT$, $N \in {\bf Z}$, we have
\begin{align} \label{13}
\MAT{x(NT) \\ p(NT)} = \CAL{L}^N \MAT{\hat{R}(\Omega (NT)/2)x \\ \hat{R}(\Omega (NT)/2)p} . 
\end{align}
Using $\zeta _1(t) \zeta _2 '(t) - \zeta _1 '(t) \zeta _2 (t) = 1$ for all $t \in {\bf R}$, we have 
\begin{align*}
\mathrm{det} ( \CAL{L} - \lambda ) = \lambda ^2 - \CAL{D} \lambda +1,  
\end{align*} 
where $\CAL{D} = \zeta _1 (T) + \zeta _2'(T)$ is referred to as the {\em discriminant of the Hill's equation}, and together with \eqref{13}, we find the following lemma:
\begin{Lem}
Let $N \in {\bf Z}$, $|N| \gg 1$. Then, for all $\phi \in C_0^{\infty} ({\bf R}^2)$, there exist constants $\lambda_e $, $C_{e}$, $C_p $, $C_{m}>0$ such that 
\begin{align*}
 (C_e)^{-1} e^{\lambda _e N}  \leq & \left\| 
x U_0(NT,0) \phi 
\right\|_{(L^2({\bf R}^2))^2} \leq C_e e^{\lambda _e N} , \quad \mbox{ if } \CAL{D}^2 >4, \\ 
 (C_p)^{-1} N  \leq & \left\| 
x U_0(NT,0) \phi 
\right\|_{(L^2({\bf R}^2))^2} \leq C_p N , \quad \mbox{ if } \CAL{D}^2 =4, \\ 
& \left\| 
x U_0(NT,0) \phi 
\right\|_{(L^2({\bf R}^2))^2} \leq C_m , \quad \mbox{ if } \CAL{D}^2 <4, 
\end{align*}
hold.
\end{Lem}
This lemma implies that for $\CAL{D}^2 =4$, the particle assumes a uniform linear motion; however, for $\CAL{D}^2 >4$, the asymptotic velocity of the particle grows exponentially in $t$. Such physical phenomena were reported by Korotyaev \cite{Ko} and the scattering theory for the case where $\CAL{D}^2 =4$ with dimensions $n=3$ and $\CAL{D}^2 >4$ with dimension $n =2,3$ have been considered. Conversely, \cite{Ka} found a relationship between the repulsive Hamiltonian $p^2-x^2$ and $\CAL{D}^2 >4$; Schr\"{o}dinger operator $p^2$ and $\CAL{D}^2 =4$, as defined by the Floquet operator. 

In this study, we focus on the case $\CAL{D}^2 >4$ and prove the asymptotic completeness of the wave operators. The scattering theory
 for a time-periodic magnetic field was first considered by \cite{Ko}, who proved the asymptotic completeness under some technical conditions of the magnetic field; thereafter, Adachi-Kawamoto \cite{AK} proved the asymptotic completeness of wave operators for the case where the magnetic field is pulsed. In the study of \cite{AK}, they discovered an explicit formula for the integral kernel of free propagators, which indicated that $\CAL{D} ^2 >4$ and $\zeta _2 (T) \neq 0$ is the best possible condition for showing asymptotic completeness in the pulsed case; if $\zeta _2 (T) = 0$, the absolute value of the integral kernel of $U_0(nT,0) \phi$, $n \in {\bf Z}$, diverges infinity for any $\phi \in L^2({\bf R}^2)$. Hence, it still remains important to determine the asymptotic completeness for {\em general time-periodic magnetic fields with only two conditions $\CAL{D} ^2 >4$ and $\zeta _2 (T) \neq 0$}; this factor is considered herein.

 \begin{Ass}\label{A1}
 Suppose that $\zeta _1 (t), \zeta _1 '(t) , \zeta _2 (t)$, and $\zeta _2 '(t)$ are continuous functions on $t \in [0,T)$, and that for all $t \in [0,T)$ and $N \in {\bf Z}$ there exist $A_{1,N}, A_{2,N}, A_{3,N}, A_{4,N} $ such that the solutions to \eqref{14} satisfy 
 \begin{align*}
 \MAT{\zeta _1 (t+ NT) \\ \zeta _2 (t +NT)} = \MAT{A_{1,N} & A_{2,N} \\ A_{3,N} & A_{4,N}} \MAT{\zeta _1(t) \\ \zeta _2 (t)}.
  \end{align*}
Moreover, for some $\lambda >0$, $\tilde{\lambda} \leq \lambda$, and for $|N| \gg 1$, there exist $0<c_3< C_3$ and $0< c_4 < C_4 $ such that 
\begin{align*}
c_3e^{\lambda N} \leq |A_{3,N} | \leq C_3 e^{\lambda N} , \quad c_4 e^{\tilde{\lambda }N} \leq |A_{4,N}| \leq C_4e^{\tilde{\lambda }N}.
\end{align*} 
holds.
 \end{Ass}
 \begin{Rem}
 Owing to Lemma 8 in \cite{Ka}, this assumption will be true for $\CAL{D}^2 = (\zeta _1 (T) + \zeta _2 '(T)) ^2 >4$ and $\zeta _2 (T) \neq 0$, and according to Kargel-Korotyaev \cite{KK}, the model of $B(t)$ such that $\zeta _2 (T) \neq 0$ is known if $B(t) $ is even. To simplify the proofs, there is need to handle this assumption. In this case, where $\CAL{D}^2 > 4$ and $\zeta _2 (T) \neq 0$, it is possible that $A_{4,N} \sim e^{- \lambda N}$ as $|N| \gg 1$ (refer also to \cite{Ko}). The assumption on $A_{4,N}$ is stated to admit such cases.
 \end{Rem}
We assume the following on the potential $V$;
\begin{Ass}\label{A2}
$V$ is a multiplication operator of $V(x)$, and $V(x) = \rho_1 (x) \rho_2 (x)$, where $\rho_1 (x) = |V(x)|^{1/2}$ and $\rho_2 (x) = \mathrm{sign} (V) |V|^{1/2}$ satisfies the following: 
$V$ is in $ C(\bfR ^2)$ and is bounded. Moreover, there exists $p>4$ such that $\rho_1, \rho_2 \in L^p ({\bf R}^2)$.
\end{Ass}  
\begin{Rem}
To simplify the proof, we used this assumption in this study. In the case where $|V(x)| \leq C \J{x}^{- \rho} $, $\rho >0$, Assumption \ref{A2} demands $\rho > 2/p$ and noting that we can consider $p>4$ to be sufficiently large, this assumption allows any small $\rho >0$. In this sense, we can consider the scattering theory even if $V$ decays slowly in $x$. 
\end{Rem}
Here, let us define $U(t,0)$ as a propagator for $H(t) = H_0(t) + V$, and the unique existence of propagator $U(t,0)$ under this assumption is guaranteed by Yajima \cite{Ya}. 
 
 Under these settings, we can define the Floquet Hamiltonian associated with $H_0$ and $H$. Let $\SCR{K}:= L^2([0,T];L^2({\bf R}^2))$ and define the self-adjoint operators acting on $\SCR{K}$ as 
 \begin{align*}
 \hat{H} = \hat{H}_0 + V, \quad \hat{H}_0 =- i\partial _t + H_0 
 \end{align*}
and term $\hat{H}$ and $\hat{H}_0$ as the Floquet Hamiltonian associated with $H_0 (t)$ and $H(t) = H_0 (t ) +V$, respectively.
Under these assumptions, we can prove the compactness of $V(\hat{H}-i)^{-1} $ on $\SCR{K}$ (refer to, e.g., \cite{Ka} (refer
also to \cite{Ko} and \cite{AK})). However, our scheme does not demand this property.
%\begin{Rem}
%If $V= V^{\mathrm{sing}}$ and there exists $R>0$, $ V^{\mathrm{sing}}(x) \equiv 0 $ for all $|x| \geq R$. Then, the relative compactness can also be proven.
%\end{Rem}

As a theorem of this study, we first obtain the following Theorem: 
\begin{Thm} \label{T3}
Under assumptions \ref{A1} and \ref{A2}, for all $\phi \in \SCR{K}$, there exists $C >0$ such that 
\begin{align*}
\sup_{\lambda \in {\bf R} \backslash \sigma_{\mathrm{pp} (\hat{H})} \, , \, \mu >0}
\left\| 
|V|^{1/2} (\hat{H} - \lambda \mp i \mu)^{-1} |V| ^{1/2} \phi 
\right\|_{\SCR{K}} \leq C \| \phi \|_{\SCR{K}}
\end{align*}
holds, where $\sigma _{\mathrm{pp}} (\hat{H})$ denotes the set of pure point spectra of $\hat{H}$.
\end{Thm}
Thanks to this theorem, the following Theorem immediately follows:
\begin{Thm}\label{T1}
Under assumptions \ref{A1} and \ref{A2}, $\sigma _{\mathrm{sing}} (\hat{H})= \emptyset$, where $\sigma _{\mathrm{sing}} (A)$ denotes the set of the singular continuous spectrum of $A$.
\end{Thm}
Owing to the correspondence of the spectrum sets between $\hat{H}$ and the monodromy operator $U(T,0)$, refer to, for example, Proposition 3.3 of M\o ller \cite{Mo}, we also have the following corollary: 
\begin{Cor}\label{C1}
Under Assumptions \ref{A1} and \ref{A2}, $\sigma _{\mathrm{sing}} (U(T,0)) =  \emptyset$.
\end{Cor}
\begin{Rem}
Recently, \cite{Ka} showed the absence of singular continuous spectra of $U(T,0)$ using the Mourre theory. However, in this study, $V \in C^2 ({\bf R}^2)$ is required, and hence, our result is successful in relaxing this condition. 
\end{Rem}
Under these assumptions, we can obtain the existence and completeness of wave operators:
\begin{Thm}\label{T2}
Under the assumption \ref{A1} and the assumption \ref{A2}, the wave operators 
\begin{align}
W^{\pm} = \mathrm{s-} \lim_{t \to \pm \infty} U(t,0)^{\ast}U_0(t,0)
\end{align}
exist and complete, that is, 
\begin{align*}
\mathrm{Ran}\left( W^{\pm} \right) = L_{\mathrm{ac}}^2(U(T,0))
\end{align*}
holds, where $L_{\mathrm{ac}}^2(U(T,0)) \subset L^2({\bf R})$ indicates the subspace of the absolutely continuous spectrum of $U(T,0)$. 
\end{Thm} 
\begin{Rem}
For the case where $V$ has singularities, by denoting $V = V^{\mathrm{r}} + V^{\mathrm{sing}}$, we can show all theorems if {\em both} $V^{\mathrm{r}}$ and $V^{\mathrm{sing}} $ are included in $L^{\tilde{p}} ({\bf R}^2)$ with some $\tilde{p} >2$. However, in this case, the potential decays slowly, and the singularity is also weak. In this sense, we do not discuss this issue.
\end{Rem}
The first approach of the proof is to show the uniform resolvent estimate (URE) for $\hat{H}_0$, in which we imitate the approach in \cite{Ko}. Owing to the URE and Kato's smooth perturbation method, Kato \cite{Kato}, we have 
\begin{align} \label{12}
\int_{{\bf R}} \left\| |V| ^{1/2} e^{-i \sigma \hat{H}_0} \phi \right\|^2_{\SCR{K}}  d \sigma \leq C \| \phi \|_{\SCR{K}}^2 .
\end{align}
Here, Korotaev's strong propagation estimate (Proposition \ref{P1}) enables us to extend URE for $\hat{H}_0$ to that of $\hat{H}$. By employing Kato's smooth perturbation method in URE for $\hat{H}$, we obtain a resolvent estimate for $\hat{H}$, and using this, we can prove the nonexistence of singular continuous spectra of $\hat{H}$. Moreover, two estimates \eqref{12} and \eqref{12} with replacement $\hat{H}_0 \to \hat{H}$ immediately prove the asymptotic completeness of the wave operators. For URE of $\hat{H}$, we employ the approach of Herbst \cite{He}. Here, we employ the strong propagation estimate (Proposition \ref{P1}). 

The key approach for characterizing the range of wave operators for time periodic systems is the {\em Howland-Yajima method}; if the wave operators $W^{\pm}$ exist and that wave operators in the sense of the Floquet Hamiltonian 
\begin{align*}
\hat{W}^{\pm} := \mathrm{s-} \lim_{\sigma \to \pm \infty} e^{i \sigma \hat{H}} e^{-i \sigma \hat{H}_0}
\end{align*}
exist and satisfy $\mathrm{Ran} \left(\hat{W}^{\pm} \right) = \SCR{K}_{\mathrm{ac}} (\hat{H})$, then the asymptotic completeness 
\begin{align*}
\mathrm{Ran} \left( W^{\pm} \right) = L_{\mathrm{ac}} (U(T,0))
\end{align*}
holds, where $\SCR{K}_{\mathrm{ac}}(\hat{H}) \subset \SCR{K}$ is the space of the absolutely continuous spectrum of $\hat{H}$. Hence, to prove Theorem \ref{T2}, we should show the existence of $W^{\pm}$, $\hat{W}^{\pm}$, and $\mathrm{Ran} \left(\hat{W}^{\pm} \right) = \SCR{K}_{\mathrm{ac}} (\hat{H})$. Here, we remark that Enss-Veseri\'{c} \cite{EV} and Kitada-Yajima \cite{KY} show the equivalence 
\begin{align*}
L^2 ({\bf R}^2) = L_{\mathrm{sc}} (U(T,0) ) \oplus L_{\mathrm{ac}} (U(T,0)) \oplus L_{\mathrm{pp}} (U(T,0)). 
\end{align*}
where $ L_{\mathrm{sc}} (U(T,0) ) $, $ L_{\mathrm{ac}} (U(T,0)) $, and $ L_{\mathrm{pp}} (U(T,0))$ denote the space of the singular continuous spectral, absolutely continuous spectral, and pure point spectra, respectively. 
\section{Uniform resolvent estimates for $\hat{H}_0$}
In this section, we consider the URE for $\hat{H}_0$ under assumption
\ref{A1}. In the following, $\| \cdot \|_p $, $1 \leq p \leq \infty$ denotes $  \| \cdot \|_{L^p({\bf R}^2)}$. The estimation key for showing the URE is the following dispersive estimates for the free propagator $U_0(t,s)$: 
\begin{Prop}\label{P1}
For all $\phi \in L^1({\bf R}^2)$, the dispersive estimates 
\begin{align} \label{1}
\left\| 
U_0(\tau,s) \phi 
\right\|_{\infty} \leq \frac{C}{ |\Gamma  (\tau,s) | } \left\| \phi \right\|_{1}
\end{align}
holds, where 
\begin{align} \label{3}
\Gamma  (\tau ,s) = \left| \zeta _1 (s) \zeta _2 (\tau) - \zeta_1(\tau) \zeta _2(s) \right|
\end{align}
Moreover, for $\eta_1, \eta _2 \in L^{\tilde{p}}({\bf R})$ with $2 \leq \tilde{p} < \infty$ and $\psi \in L^2({\bf R}^2)$, 
\begin{align} \label{2}
\left\| 
\eta_1 U_0(\tau ,s) \eta_2 \psi 
\right\|_2 \leq C |\Gamma (\tau ,s)|^{-2/\tilde{ p} } \left\| \eta_1 \right\|_{\tilde{p}} \left\| \eta_2 \right\|_{\tilde{ p} } \left\| \psi  \right\|_2.
\end{align}
\end{Prop}
\Proof{
This proposition was proven by \cite{Ko} for a special case of magnetic fields. Inequality \eqref{1} can be shown using (35) in \cite{Ka} (or refer also to (7.7) of \cite{AK} and Lemma 2.3 in Kawamoto \cite{Ka2}). Hence, we only show \eqref{2}. Owing to the Riesz-Thorin interpolation theorem, we have for $2 \leq Q \leq \infty$ and $\phi \in L^{Q/(Q-1)} ({\bf R}^2)$
\begin{align*} 
\left\| 
U_0(\tau,s) \phi
\right\|_Q \leq C |\Gamma (\tau, s)|^{-2(1/2 - 1/Q)} \left\| \phi \right\|_{(Q/(Q-1))}
\end{align*}
holds. Hence, for $\psi \in L^2({\bf R}^2)$ and $Q = 2\tilde{ p} /( \tilde{ p} -2)$, 
\begin{align}
\nn \left\| 
\eta_1 U_0(\tau,s) \eta_2 \psi
\right\|_2 & \leq C \left\| \eta_1 \right\|_{\tilde{p}} \left\| U_0(\tau,s) \eta_2 \psi\right\|_Q \\  
\label{20} & \leq   C |\Gamma (\tau ,s)|^{-2/\tilde{ p} } \left\| \eta_1 \right\|_{\tilde{p}} \left\| \eta_2 \psi\right\|_{2\tilde{ p} /( \tilde{ p} +2)} 
\\  
\nn & \leq   C |\Gamma (\tau ,s)|^{-2/ \tilde{ p} } \left\| \eta_1 \right\|_{\tilde{p} } \left\| \eta_2 \right\|_{\tilde{p} } \left\| \psi  \right\|_2.
\end{align}
}

Before we show the URE for $\hat{H}_0$, we remark some properties for $\zeta _j (t)$, $j=1,2$. By the definition of $\zeta _j (t)$, one discovery 
\begin{align} \label{17}
\zeta _1 (t) \zeta _2 '(t) - \zeta _1 '(t) \zeta _2 (t) = 1.
\end{align}
Owing to this equation, one sees that zero points for $\zeta _j (t)$ and $t \in [0,T)$ are unique. Indeed, if $t _0 \in [0, T)$ exists such that $\zeta _1 (t_0) = 0$, then \eqref{17} yields $ \zeta _1 '(t_0) \zeta _2 (t_0) = -1 \neq 0$, that is, $\zeta _1 '(t_0)  \neq 0$. The same is true for $\zeta _2 (t)$. Now, we divide $[0, T)$ into 
\begin{align} \label{18}
[0,T) = \Omega _1^l \cup \Omega _2^l \cup \Omega _{J_l}^l \cup \{0, t_1^{(l)}, t_2^{(l)}, ..., t_{J_l-1}^{(l)} \}, \quad J_l \in {\bf N}, \quad  l =1,2
\end{align}
where $\zeta _l (t_k^{(l)}) = 0$, $k \in \{ 1,2,...,J_{l}-1 \}$, and $\zeta _l (t) \neq 0$ for all $t \in \Omega _j^l$, $j \in \{ 1,..,J_l \}$. Noting \eqref{17}, we also notice that $\zeta _1 (t) / \zeta _2 (t)$ (resp. $\zeta _2 (t) / \zeta _1 (t)$) is a monotonically increasing function (resp. the monotone decreasing function) on $\Omega ^2 _{j}$ (resp. $\Omega ^1 _j$) and satisfies 
\begin{align*}
\frac{d}{dt} \frac{\zeta _1 (t)}{\zeta _2 (t)} = \frac{1}{(\zeta _2 (t) )^2} >0, \quad \left( 
\mbox{resp.} \quad \frac{d}{dt} \frac{\zeta _2 (t)}{\zeta _1 (t)} = - \frac{1}{(\zeta _1 (t))^2} <0
\right).
\end{align*}

\begin{Prop} \label{P2}
For all $\phi \in \SCR{K}$, there exists a constant $C>0$ such that
\begin{align*}
\sup_{\lambda \in {\bf R}, \, \mu >0} \left\| 
\rho_1 (\hat{H}_0 -\lambda \mp i \mu)^{-1} \rho_2  \phi
\right\|_{\SCR{K}} \leq C \| \phi \|_{\SCR{K}}
\end{align*}
\end{Prop}
\Proof{
The fundamental proof is based on
\cite{Ko}. Owing to the Laplace transform, we have: 
\begin{align} \label{27}
 (\rho_1 (\hat{H}_0 - \lambda - i \mu)^{-1} \rho _2
 \phi) (t,x) = -i \rho_1 (x)  \int_0^{\infty} \left( e^{- i \sigma (\hat{H}_0 - \lambda - i\mu ) }  \rho _2
 \phi \right) (t,x) d \sigma. 
\end{align} 
Then, using the formula of $e^{- i\sigma \hat{H}_0}$, consider, for example, Yajima \cite{Ya2}, we have 
\begin{align*}
(\rho_1 (\hat{H}_0 - \lambda - i \mu)^{-1} \rho _2
 \phi) (t,x) &= i \rho_1 (x) \sum_{N=1}^{\infty} \int_0 ^T e^{i(t+ NT -s) (\lambda + i \mu) } U_0(t+NT, s ) (\rho _2
 \phi) (s,x) ds \\ & \quad + i \rho_1 (x) \int_0^t e^{i(t-s) (\lambda + i\mu)} U_0(t,s) (\rho _2
 \phi) (s,x) ds. 
\end{align*} 
Then, for $p  > 4$, $\lambda \in {\bf R}$ and $\mu >0$, using Proposition \ref{P1}, we have 
\begin{align}
\nn  &\| (\rho_1 (\hat{H}_0 - \lambda \mp i \mu)^{-1} \rho _2
 \phi) \|_{\SCR{K}} 
\\ \nn  & \leq C \sum_{N=0}^{\infty}
\left\|
\int_0^T \|  (\rho_1 U_0(t+NT,s) \rho _2
 \phi(s))\|_{2} ds 
\right\|_{L^2([0,T])} \\
\nn  & \leq 
C\sum_{N=0}^{\infty} \|\rho_1\|_{p} \| \rho _2\|_{p} 
\left\|
\int_0^T |\Gamma (t+NT , s)|^{-2/p} \| \phi(s)\|_{2} ds 
\right\|_{L^2([0,T])} \\
& \leq \label{4} 
C\sum_{N=0}^{\infty} \|\rho_1\|_{p} \| \rho _2\|_{p} \| \phi\|_{\SCR{K}} \left(
\int_{[0,T]^2} |\Gamma (t+NT ,s)|^{-4/ p} dsdt
\right)^{1/2} .
\end{align}
By \eqref{3}, 
\begin{align*}
& \int_0^T \int_0^T |\Gamma  (t+NT,s)|^{-4/p} ds dt \\ & =
 \int_0^T |\zeta _2 (t+NT)|^{-4/p} \int_0^T
|\zeta _2 (s)| ^{-4/p} |(\zeta _1(s) / \zeta _2(s) - \alpha)|^{-4/p} ds dt
\end{align*}
holds, where $\alpha = \alpha (t) = \zeta _1(t+NT)/\zeta _2(t+NT)$. We decompose  
\begin{align*}
 \int_0^T
|\zeta _2 (s)| ^{-4/p} |(\zeta _1(s) / \zeta _2(s) - \alpha)|^{-4/p} ds = \sum_{J=1}^{J_2} \int_{\Omega_J^2} |\zeta _2 (s)| ^{-4/p} |(\zeta _1(s) / \zeta _2(s) - \alpha)|^{-4/p} ds, 
\end{align*} 
where $J_2$ and $\Omega_J^2$ are defined by the same rules as in \eqref{18}. Here, we remark that $J_2$ can be considered a finite integer. Then, by denoting $\tau =
\zeta _1(s)/\zeta _2(s)$ and employing $d\tau/ds = (\zeta _2(s) ) ^{-2}$, we have 
\begin{align*}
\int_{\Omega_J^2}
|\zeta _2 (s)| ^{-4/p} |(\zeta _1(s) / \zeta _2(s) - \alpha)|^{-4/p} ds \leq  I_1 + I_2 + I_3
\end{align*}
with 
\begin{align*}
I_1  &: = \int_{|\tau - \alpha| \leq 1} |\zeta _2 (s)| ^{2-4/p} |\tau - \alpha|^{-4/p}
d \tau \leq 
C\int_{|\tau - \alpha| \leq 1}  |\tau - \alpha|^{-4/p}
d \tau 
\leq C , \\ 
I_2 &:=  \int_{  |\tau - \alpha| \geq 1,  \ |\tau| \leq 1 } |\zeta _2 (s)| ^{2-4/p} |\tau - \alpha|^{-4/p}
d \tau 
\leq C   \int_{  |\tau| \leq 1 } 
d \tau
\leq C
\end{align*}
and 
\begin{align*}
I_3 := \int_{|\tau - \alpha| \geq 1, \ |\tau| \geq 1}  |\zeta _2 (s)| ^{2-4/p} |\tau - \alpha|^{-4/p}
d \tau
\end{align*}
where we use $p > 4$. Furthermore, $|\zeta _2(s) |^{-1} \leq | \tau |^{-1} |\zeta _1(s)| \leq
C |\tau|^{-1}$, we also have 
\begin{align*}
 I_3  \leq C \int_{|\tau - \alpha| \geq 1, \ |\tau| \geq 1}  |\tau| ^{-2+4/p} |\tau - \alpha|^{-4/p}
d \tau \leq C
\end{align*}
by $p>4$. Combining all, one can obtain a constant $C>0$, independent of $t$, such that
\begin{align*}
\sum_{J=1}^{J_2} \int_{\Omega_J^2} |\zeta _2 (s)| ^{-4/p} |(\zeta _1(s) / \zeta _2(s) - \alpha)|^{-4/p} ds \leq C.
\end{align*}
Hence, one has 
\begin{align} \label{25}
\| (\rho_1 (\hat{H}_0 - \lambda \mp i \mu)^{-1} \rho _2
 \phi) \|_{\SCR{K}}  \leq C  \left\| |V| ^{1/2} \right\|_{p}^2 \| \phi\|_{\SCR{K}}  \sum_{N=0}^{\infty} \left( \int_0^T |\zeta _2 (t+NT)|^{-4/p} dt \right)^{1/2} .
\end{align}

We now calculate the integral in $t$. By $\zeta _2(t+NT) = A_{3,N} \zeta
 _1(t) +A_{4,N} \zeta _2(t)$, we have 
\begin{align} \label{22}
\int_0^T |\zeta _2 (t+NT)| ^{-4/p} dt \leq 
|A_{3,N}|^{-4/p} \int_0^T |\zeta _1 (t)|^{-4/p} \left| 
1 + r_0 (\zeta _2(t)/ \zeta _1(t))
\right|^{-4/p} dt
\end{align}  
where $r_0 = A_{4,N} / A_{3,N}$. We again decompose  
\begin{align*}
 \int_0^T |\zeta _1 (t)|^{-4/p} \left| 
1 + r_0 (\zeta _2(t)/ \zeta _1(t))
\right|^{-4/p} dt = \sum_{J=1}^{J_1} \int_{\Omega_J^1} |\zeta _1 (t)|^{-4/p} \left| 
1 + r_0 (\zeta _2(t)/ \zeta _1(t))
\right|^{-4/p} dt, 
\end{align*} 
where $J_1$ and $\Omega_J^1$ are defined by the same rules as in \eqref{18}. By denoting $\sigma =\zeta _2(t)/
\zeta _1(t) $, we have $d \sigma /(dt) = -( \zeta _1(t))^{-2}$ and 
\begin{align*}
\int_{\Omega _J^1} |\zeta _1 (t)|^{-4/p} \left| 
1 + r_0 (\zeta _2(t)/ \zeta _1(t))
\right|^{-4/p} dt \leq  I_4 + I_5 + I_6
\end{align*}
with 
\begin{align*}
I_4 :=& \int_{|1+r_0 \sigma| \geq 1/2 , \ |\sigma| \leq 1} 
 |\zeta _1 (t)|^{2-4/p} \left| 
1 + r_0 \sigma
\right|^{-4/p} d \sigma \leq C, \\ 
I_5 := & \int_{|1+r_0 \sigma| \geq 1/2 , \ |\sigma| \geq 1} 
 |\sigma|^{-2+4/p} \left| 
1 + r_0 \sigma
\right|^{-4/p} d \sigma \leq C.
\end{align*} 
and 
\begin{align*}
I_6 := \int_{|1 + r_0 \sigma| \leq 1/2} |\zeta _1(t)|^{2-4/p} |1+ r_0
 \sigma|^{-4/p} d \sigma .
\end{align*}
Now, we estimate $I_6$. Based on assumption \ref{A1}, there are $\lambda >0$ and $\tilde{\lambda} \leq \lambda$ such that 
$$ (c_4/C_3) e^{-( \lambda - \tilde{\lambda} ) N} \leq | r_ 0 | \leq (C_4/c_3) e^{-( \lambda - \tilde{\lambda} ) N}. 
$$ holds. Then, it can be calculated that $| \sigma | \geq C e^{ (\lambda - \tilde{\lambda} ) N}$, that is, $|\sigma|^{-1} \leq C e^{- (\lambda - \tilde{\lambda} ) N}$ on the support of $|1 + r_0 \sigma| \leq 1/2$, which yields $|\zeta _1 (t)| \leq |\sigma|^{-1} |\zeta _2 (t)| \leq C e^{- (\lambda - \tilde{\lambda}) N} $. Thus, we also have 
\begin{align*}
I_6 &\leq C e^{-N (\lambda - \tilde{\lambda} ) (2-4/p)} \int_{|1 + r_0 \sigma| \leq 1/2}|1+ r_0
 \sigma|^{-4/p} d \sigma  \\ & \leq Ce^{-N (\lambda - \tilde{\lambda}) (2-4/p)} |r_0|^{-1}
 \\ & \leq Ce^{- N( \lambda - \tilde{\lambda} ) (1-4/p)} \\ & \leq C
 .
\end{align*}
Thus, we finally obtain that for some $C > 0$, 
\begin{align} \label{26}
\int_{0}^T |\zeta _2 (t+NT)|^{-4/p} dt \leq Ce^{-4 \lambda N /p}
\end{align}
holds for $p > 4$. Finally, from \eqref{4}, \eqref{25}, and \eqref{26}, we obtain 
\begin{align*}
\left\| \rho_1 (\hat{H}_0- \lambda \mp i \mu)^{-1} \rho_2 \phi \right\|_{\SCR{K}} \leq C \left\| |V|^{1/2} \right\|_{p}^2 \left\| \phi \right\|_{\SCR{K}} \sum_{N=0}^{\infty}e^{-2 \lambda N /p}  & \leq C \| \phi \|_{\SCR{K}}
\end{align*}
It proves Proposition \ref{P2}.

}

\section{Uniform resolvent estimate for $\hat{H}$}

In this section, we present the URE for $\hat{H}$. To demonstrate this, we employ the approach according to Herbst \cite{He}. However, in \cite{He}, among the specific conditions, only the Stark Hamiltonian has been fully used, and imitating this approach may be difficult. To overcome this difficulty, we should find the alternative condition of $\hat{H}$, as well as the following {\em Lipschitz continuity for resolvent of $\hat{H}_0$}, which plays a crucial role in mimicking the approach of \cite{He}:  

\begin{Thm}\label{T5}
Let $z _{\pm},w_{\pm} \in {\bf C}_{\pm}$. Then, for all $\phi \in \SCR{K}$, 
\begin{align*} 
\left\| \left( \rho_1 (\hat{H}_0 -z_{\pm} )^{-1 } \rho_2 - \rho_1 (\hat{H}_0 -w_{\pm})^{-1} \rho_2
\right) \phi \right\|_{\SCR{K}} \leq C |z_{ \pm} - w_{\pm}| \| \phi \|_{\SCR{K}}. 
\end{align*}
holds.
\end{Thm}
\begin{Rem}
In the case where $k_0 = - \Delta$, the H\"{o}lder continuity 
\begin{align}\label{10}
\left\|  \J{x}^{-s} \left(  ({k}_0 -z_{\pm} )^{-1 }  -  ({k}_0 -w_{\pm})^{-1}\right) \J{x}^{-s}
\phi \right\|_{2} \leq C |z_{ \pm} - w_{\pm}|^{(2s-1)/(2s +1)} \| \phi \|_{2}
\end{align}
holds. Here, formally, we consider $s \to \infty$; then, the power on $|z_{\pm} -w_{\pm}|$ in \eqref{10} tends to $1$ (that is, the Lipschitz continuity). In our model, compared with the exponential growth of $|x(t)|$ in $t$, the decay for the potential $\rho_j \in L^p({\bf R}^2)$ is fast. In the words of $k_0$, the potential decays exponentially in $x$. Thus, it is not surprising that Lipschitz continuation \eqref{10} holds.  
\end{Rem}
This theorem can be obtained as the sub-consequence of the following lemma;
\begin{Lem}\label{L4} Suppose Assumption \ref{A1}. For a sufficiently small $\ep >0$, let $\tau \in [- \ep ,
 \ep]$. Define  
\begin{align*}
\Sigma _{R} = \int_0^R \left\|
\sigma \rho _1 e^{-i (H_0 + \lambda \pm \tau i) \sigma}(\rho _2 \phi)
\right\|_{\SCR{K}} d \sigma
\end{align*}
for $R >0$. Then, there exists $\ep >0$ such that for all $\tau \in [- \ep, \ep]$ and $\lambda \in {\bf R}$, the limit $\lim_{R \to \infty} \Sigma _R $ exists, satisfying
\begin{align*}
\lim_{R \to \infty} \Sigma _R < C \| \phi \|_{\SCR{K}} .
\end{align*}
Furthermore,
\begin{align*}%\label{300}
\mathrm{s-}\lim_{\sigma \to \infty} \sigma \rho _1 e^{-i (H_0 + \lambda
 \pm \tau i) \sigma} \rho _2 = 0
\end{align*} 
holds.
\end{Lem}
\Proof{
Let $R = N_0T + s$ with $N_0 \in \bfN$ and $s \in [0,T]$. Then, the same calculations in the proof of Proposition \ref{P2} yield
\begin{align*}
\Sigma _R &\leq C \sum_{N=0}^{N_0 +1} \int_0^T
NTe^{\ep NT} \| \rho_1 U_0(t+NT , s) (\rho_2
 \phi)\|_{\SCR{K}} ds \\ 
& \leq  
C \sum_{N=0}^{N_0 +1} |NT|e^{\ep NT} e^{-2\lambda N /p} \| \phi\|_{\SCR{K}}
\end{align*}
with $p >4$. Then, if $|\ep|$ is sufficiently small, there exists $\ep _1 > 0$ such that 
$ (|NT|e^{\ep NT} e^{-2 \lambda /p}) \leq C e^{- \ep _1
N}$. This implies that $\lim \Sigma _R $ exists and $\lim \Sigma _R < \infty$.}

Because 
\begin{align*}
   \rho_1 (\hat{H}_0 -z_{\pm} )^{-1 } \rho_2 \phi- \rho_1 (\hat{H}_0 -w_{\pm})^{-1} \rho_2 \phi
  &= 
\mp i \rho_1 \int_0^{\infty} \left( e^{\mp i \sigma (\hat{H}_0 - z_{\pm}) } - e^{\mp i \sigma (\hat{H}_0 -w_{\pm})} \right) \rho_2 \phi d \sigma \\ 
&= 
\mp i \int_0^{\infty} \sigma \rho_1  e^{\mp i \sigma \hat{H}_0} \rho_2  \left( \frac{e^{\mp i \sigma z_{\pm}} - e^{\mp i \sigma w_{\pm}}}{\sigma} \right)\phi d \sigma, 
\end{align*}
and Lemma \ref{L4} with $\lambda = \tau =0$, Theorem \ref{T5} can be proven.

\subsection{Proof of Theorem \ref{T3}}
Now, we prove Theorem \ref{T3}. The fundamental approach is based on the approach described by Herbst \cite{He}. 
By the resolvent formula, we have that
\begin{align*}
\rho_1 (\hat{H} - \lambda - i \ep)^{-1}\rho_2 = 
(1+ \rho_1  (\hat{H} - \lambda - i \ep)^{-1}\rho_2)^{-1} \cdot \rho_1 (\hat{H}_0 - \lambda - i \ep)^{-1}\rho_2
\end{align*}
holds. Hence, we prove that for any $\ep \geq 0$, if $\psi  = \psi_{\ep} \in \SCR{K}$ such that 
$
(1+ \rho_1  (\hat{H} - \lambda - i \ep)^{-1}\rho_2) \psi = 0, 
$, then $\psi \equiv 0$. Therefore, we have that $(1+ \rho_1  (\hat{H} - \lambda - i \ep)^{-1}\rho_2)$ is invertible for any $\ep \geq 0$, which proves that Theorem \ref{T3} holds. 

Let $ \varphi  = (\hat{H}_0 - \lambda -i
\ep)^{-1} \rho_2 \psi$, then 
\begin{align} \label{9}
(\hat{H} - \lambda -i \ep) \varphi = 0
\end{align}
holds. Here, remarking $\hat{H}$ is a selfadjoint operator and $
\lambda \not \in \sigma _{pp}(\hat{H})$, \eqref{9} implies $ \varphi \equiv 0$, (i.e., $\psi \equiv 0$) if 
$\varphi \in \SCR{D}(\hat{H})$. Thus, we prove $\varphi \in \SCR{D}
(\hat{H})$, as follows: Using the resolvent formula 
\begin{align*}
\varphi = (\hat{H}_0 -i)^{-1} \rho_2 \psi + (\lambda + i \ep
 -i)(\hat{H}_0 -i )^{-1} \varphi,
\end{align*}
we notice that it suffices to show that $\varphi \in \SCR{K}$ shows $\varphi \in \D{\hat{H}}$. Hence, we show $\varphi \in \SCR{K}$. In the case where $\ep >0$, it is clear that $\varphi \in \SCR{K}$; hence, we only consider the case where $\ep \to 0$, that is, $\psi = -\rho_1 (\hat{H}_0 - \lambda -i 0)^{-1} \rho_2 \psi$ and $\varphi = (\hat{H}_0 - \lambda -i 0)^{-1} \rho_2 \psi$:
\begin{align*}
\left\| 
\varphi
\right\|_{\SCR{K}} ^2 
&= \lim_{\ep \to 0}\left\| (\hat{H}_0 - \lambda -i \ep)^{-1} \rho_2 \psi  \right\|^2_{\SCR{K}} 
\\ &= \lim_{\ep \to 0} \left\{ \left( \rho_2 (\hat{H}_0 - \lambda + i \ep)^{-1} (\hat{H}_0 - \lambda -i \ep)^{-1} \rho_2 \psi, \psi \right)  \right\}
\\ &= \lim_{\ep \to 0} \left\{  \frac{-i}{2 \ep} \left( \rho_2 \left( (\hat{H}_0 - \lambda -i \ep)^{-1} -  (\hat{H}_0 - \lambda + i \ep)^{-1} \right) \rho_2 \psi, \psi \right) \right\}
\\ &=  \lim_{\ep \to 0} \left\{  \frac{-i}{2 \ep} \left( \rho_2 \left( (\hat{H}_0 - \lambda -i \ep)^{-1} -  (\hat{H}_0 - \lambda + i \ep)^{-1} \right) \rho_2 \psi, \psi \right) - \frac{i}{2 \ep} \left(  (\psi, \psi) - (\psi, \psi) \right) \right\}
\\ &= \lim_{\ep \to 0} \lim_{\delta _1 \to 0} \left\{ \frac{-i}{2 \ep} \left( \left( \rho_2 (\hat{H}_0 - \lambda -i \ep)^{-1} \rho_2 - \rho_1  (\hat{H}_0 - \lambda - i \delta_1)^{-1} \rho_2 \right)\psi, \psi \right) \right\} \\ & \quad - 
 \lim_{\ep \to 0} \lim_{\delta _2 \to 0} \left\{ \frac{-i}{2 \ep} \left( \psi, \left( \rho_2 (\hat{H}_0 - \lambda -i \ep)^{-1} \rho_2 - \rho_1  (\hat{H}_0 - \lambda - i \delta_2)^{-1} \rho_2 \right)\psi \right) \right\}
\\ &=: \lim_{\ep \to 0} \frac{-i}{2 \ep} \left(  I_1(\ep) -  I_2(\ep) \right).
\end{align*}
We have 
\begin{align*}
I_{1} (\ep) = \lim_{\delta _1 \to 0} \left( \left( \rho_2 \left( (\hat{H}_0 - \lambda -i \ep)^{-1} - (\hat{H}_0 - \lambda - i \delta_1)^{-1} \right) \rho_2 \right)\psi, \psi \right) + \left( \frac{\rho_1 - \rho_2}{\rho_1} \psi, \psi \right).
\end{align*}
and 
\begin{align*}
I_{2} (\ep) = \lim_{\delta _2 \to 0} \left( \psi, \left( \rho_2 \left( (\hat{H}_0 - \lambda -i \ep)^{-1} - (\hat{H}_0 - \lambda - i \delta_2)^{-1} \right) \rho_2 \right) \psi \right) + \left(  \psi, \frac{\rho_1 - \rho_2}{\rho_1} \psi \right).
\end{align*}
These and \eqref{10} yield 
\begin{align*}
\left\| \varphi \right\|_{\SCR{K}} ^2 \leq C \left(  \lim_{\ep \to 0}\lim_{\delta _1 \to 0} \frac{|\ep - \delta _1|}{\ep} +   \lim_{\ep \to 0}\lim_{\delta _2 \to 0} \frac{|\ep - \delta _2|}{\ep} \right) \| \psi \|_{\SCR{K}}^2 \leq C \| \psi \|_{\SCR{K}}^2, 
\end{align*}
where we use 
\begin{align*}
\left( \frac{\rho_1 - \rho_2}{\rho_1} \psi, \psi \right)- \left(  \psi, \frac{\rho_1 - \rho_2}{\rho_1} \psi \right) = 0.
\end{align*}
Therefore, we obtain $\varphi \equiv 0$ and which implies $\psi \equiv 0$.

\subsection{Proof of Theorem \ref{T1}}

Next, we show the absence of a singular continuous spectrum of $\hat{H}$. In the case where $V \in C^2({\bf R}^2)$, \cite{Ka} showed this issue using the Mourre inequality. Hence, we relaxed this condition. From Theorem \ref{T3}, we have that 
\begin{align*}
\sup_{\lambda \in {\bf R} \backslash \sigma_{\mathrm{pp}} (\hat{H}) , \ep >0} \left\| 
|V|^{1/2} \left( (\hat{H} - \lambda -i\ep)^{-1} - (\hat{H} - \lambda + i \ep)^{-1}  \right) |V|^{1/2} \phi
\right\|_{\SCR{K}} \leq C \left\|  \phi \right\|_{\SCR{K}}. 
\end{align*}
Here, we let $(\cdot , \cdot)$ be an inner product of $L^2({\bf R}^2)$. Then, for all $\lambda _1, \lambda _2 \in {\bf R}$, 
\begin{align*}
& \left| \left( (E_{\hat{H}} (\lambda_1) - E_{\hat{H}} (\lambda _2) )|V|^{1/2} \phi , |V|^{1/2} \phi \right) \right|.
\\ & = \left| \lim_{\ep \to 0} \frac{1}{2 \pi i} \int_{\lambda _2 }^{\lambda _1 } \left( |V|^{1/2} \left( (\tau - \lambda -i\ep)^{-1} - (\tau - \lambda + i \ep)^{-1}  \right) |V|^{1/2} \phi , \phi  \right)d \tau \right| \\ & \leq C |\lambda _2 - \lambda _1| \left\| \phi \right\|_{\SCR{K}} ^2
\end{align*}
holds, where $E_{\hat{H}} (\cdot)$ is the spectral decomposition of $\hat{H}$, implying that for the connected Borel set $I := (\lambda _1, \lambda _2) \subset {\bf R} \backslash \sigma_{\mathrm{pp}} (\hat{H})$, $E (I) |V|^{1/2} \phi \in \SCR{K}_{\mathrm{ac}} (\hat{H})$. Here, $|V|^{1/2}$ is the closed operator, which yields for all $u \in \SCR{K}$, $E(I) u \in \SCR{K}_{\mathrm{ac}}(\hat{H})$. Because we can arbitrarily take $\lambda _1, \lambda _2$, we find $\SCR{K} = \SCR{K}_{\mathrm{ac}} (\hat{H})$, that is, $\SCR{K}_{\mathrm{sing}} (\hat{H}) = \emptyset $.

\section{Asymptotic completeness}

Finally, the main theorem is presented. By Proposition \ref{P2}, Theorem \ref{T3} and Kato's smooth perturbation method, we get for all $\varphi \in C_0^{\infty} ({\bf R} \backslash \sigma_{\mathrm{pp}} (\hat{H}) ) $ and $\phi, \psi \in \SCR{K}$, 
\begin{align} \label{7}
\int_{- \infty}^{\infty} \left\| |V|^{1/2} \varphi(\hat{H}) e^{-i \sigma \hat{H}} \phi \right\|_{\SCR{K}}^2 d \sigma \leq C \| \varphi(\hat{H}) \phi \|_{\SCR{K}} ^2 , \quad 
\int_{- \infty}^{\infty} \left\| |V| ^{1/2} e^{-i \sigma \hat{H}_0} \psi \right\|_{\SCR{K}}^2 d \sigma \leq C \|  \psi \|_{\SCR{K}} ^2
\end{align} 
holds. Hence for all $\sigma_1, \sigma_2 \in {\bf R}_{\pm} := \left\{ a \in {\bf R} \, | \, \pm a \geq 0 \right\}$, 
\begin{align*}
& \left\| 
e^{i \sigma _1 \hat{H}_0} e^{-i \sigma _1 \hat{H} } \varphi(\hat{H})\phi  - e^{i \sigma _2 \hat{H}_0} e^{-i \sigma _2 \hat{H} } \varphi(\hat{H})\phi
\right\|_{\SCR{K}} 
\\ & \leq \sup_{\| \psi \|_{\SCR{K} =1}}\int_{\sigma _2}^{\sigma _1} \left| \left( 
\rho_1 e^{-i \sigma \hat{H}} \varphi(\hat{H})\phi , \rho_2 e^{-i \sigma  \hat{H}_0} \psi
\right) \right| d \sigma \\ & \leq 
\sup_{\| \psi \|_{\SCR{K} =1}} \left( \int_{\sigma _2}^{\sigma _1} \left\| |V|^{1/2} e^{-i \sigma \hat{H}} \varphi(\hat{H}) \phi \right\|_{\SCR{K}} ^2 d \sigma \right)^{1/2}
\left( \int_{\sigma _2}^{\sigma _1} \left\| |V|^{1/2} e^{-i \sigma \hat{H}_0} \psi \right\|_{\SCR{K}} ^2 d \sigma \right)^{1/2}
\\ & \to 0, \quad \mbox{as $\sigma _1, \sigma _2 \to \infty$}
\end{align*} 
holds, where we use \eqref{7}. Hence, there is a strong limit: $ \mathrm{s-}\lim_{\sigma \to  \pm \infty} e^{i \sigma  \hat{H}_0} e^{-i \sigma  \hat{H} } \varphi(\hat{H})$ exists for all $\varphi \in C_0^{\infty} ({\bf R} \backslash \sigma_{\mathrm{pp}} (\hat{H}))$. Similarly, we have a strong limit: $ \mathrm{s-} \lim_{\sigma \to \pm \infty} e^{i \sigma  \hat{H}} e^{-i \sigma  \hat{H}_0 }$. Consequently, we obtain 
$$\mathrm{Ran} \left(  \mathrm{s-} \lim_{\sigma \to \pm \infty} e^{i \sigma  \hat{H}} e^{-i \sigma  \hat{H}_0 } \right) = \SCR{K} _{\mathrm{ac}} (\hat{H}).$$ Using the Howland-Yajima method, the proof for Theorem \ref{T3} is completed by showing the existence of $W^{\pm}$; more precisely, refer to \S{4} of \cite{Ya2} or the end \S{6} of \cite{AK}. We now show the existence of $W^{\pm}$. Owing to the Cook-Kuroda method, it suffices to show that for all $\phi \in \SCR{S} ({\bf R}^2)$, 
\begin{align} \label{19}
& \int_1^{\infty} \left\| |V|^{1/2} U_0(t,0) \phi \right\|_{L^2({\bf R}^2)} dt \leq C. 
\end{align}
Hence, we show that
\begin{align} \label{21}
 \sum_{N=1}^{\infty} \int_0^{T} \left\| |V|^{1/2} U_0(t+NT,0) \phi \right\|_{2} dt \leq C .
\end{align}
holds. Then, \eqref{19} holds. By noting \eqref{20}, for $p>4$, the left hand side of \eqref{21} is smaller than 
\begin{align*}
C \left\| |V|^{1/2} \right\|_p \| \phi \|_{2p/(p+2)} \sum_{N=0}^{\infty} \int_0^{T} |\zeta _2 (t+NT)|^{-2/p} dt, 
\end{align*}
where we also use \eqref{1} and \eqref{2} with $\tau =t+NT$ and $s=0$. From \eqref{26}, we obtain \eqref{21}.


\begin{thebibliography}{9999}

\bibitem{AK} Adachi, T., Kawamoto, M.: Quantum scattering theory in a periodically pulsed magnetic field, Annales Henri Poincar\'{e}, \textbf{17}, 2409--2438, (2016). 

\bibitem{EV} Enss, V., Veseli\'{c}, K.: Bound states and propagating states for time-dependent hamiltonians, Ann., de l'I.H.P, section A, \textbf{39}, 159--191, (1983). 

\bibitem{He} Herbst, I.W.: Unitary equivalence of Stark Hamiltonians,
	    Math. Z., \textbf{155}, 55--70, (1977) .   
%\bibitem{BCHM} J.F.Bony, R.Carles D.H\"{a}fner and L.Michel.: Scattering theory for the Schr\"{o}dinger equation with repulsive potential. (2009) 
%\bibitem{CFKS} Cycon, H. L., Froese, R. G., Kirsch, W., Simon, B.:
%	      Schr\"{o}dinger operators, Text and Monographs in
%	      Physics. Springer-Verlag (1987) 
%\bibitem{CS} Colombini, F., Spagnolo, S.: Hyperbolic equation with
%	    coefficients rapidly oscillating in time: A result of
%	    nonstability, J.D.E. \textbf{52}, 24-38 (1984)   
%\bibitem{DG} Derezi\'{n}ski, J., G\'{e}rard, C.: Scattering theory
%	    of classical and quantum N-particle systems, Text
%	    Monogr. Phys., Springer, Berlin, 1997
%\bibitem{GL} G\'{e}rard, C., \L aba, I.: Multiparticle quantum
%	     scattering in constant magnetic fields, In: Mathematical
%	     surveys and Monographs,
%	     \textbf{90}. American. Mathematical. Society, Providence,
%	     RI (2002)   
%\bibitem{HS} Helffer, B., Sj\"{o}strand, J.: Equation de
%	    Schr\"{o}dinger avec champ magn\'{e}tique et equation de
%	    Harper, Springer Lecture Notes in Physics \textbf{345}
%	    118-197 (1989) 

%\bibitem{Ho} Hodchstadt, H.: Functiontheoretic properties of the
%	     discriminant of Hill's equation, Math. Z. \textbf{82},
%	     237-242 (1963)
\bibitem{How} Howland, J.S.: Scattering theory for hamiltonians
	     periodic in time, Indiana Univ. Math. J. \textbf{28},
	     471--494, (1979).   
	     
\bibitem{KK} Kargaev, P., Korotyaev, E.L.: The inverse problem for the Hill operator, a direct approach. Inv. Math., \textbf{129}, 567--593, (1997).  

\bibitem{Kato} Kato, T.: Wave operators and similarity for some
	    non-self-adjoint operators, math. Ann. \textbf{162}, 258--279,
	    (1966).   
	    
\bibitem{Ka} Kawamoto, M.: Mourre theory for time-periodic magnetic fields, J. Func., Anal., \textbf{277}, 1--30, (2019). 

\bibitem{Ka2} Kawamoto, M.: Strichartz estimates for Schr\"{o}dinger operators with square potential with time-dependent coefficients, Differ. Equ. Dyn. Syst., (2020). 
%\bibitem{KaKu} Kato, T., Kuroda, S. T.: The abstract theory of
%	      scattering, Rockey Mountain J. Math. \textbf{1}, 127-171
%	      (1971)  
\bibitem{KY} Kitada, H., Yajima, K.: Bound states and scattering states for time periodic hamiltonians, Ann., de l'I.H.P, section A, \textbf{39}, 145--157, (1983).

\bibitem{Ko} Korotyaev, E. L.: On scattering in an external,
	    homogeneous, time-periodic magnetic field,
	    Math. USSR-Sb. \textbf{66}, 499--522 (1990) 
%\bibitem{MW} Magnus and Winkler.: Hill's Equations, Tracts in
%	    Mathematics \textbf{20}, (1966)
\bibitem{Mo} M\o ller, J. S.: Two-body short-range systems in a
	    time-periodic electric field, Duke Math. J. \textbf{105}
	    135-166 (2000). 
%\bibitem{Mo} Mourre, E.: Absence of singular continuous spectrum for
%	    certain selfadjoint operators,
%	    Comm. Math. Phys. \textbf{91} 391-408 (1981) 
%\bibitem{X} Rongcong Xu.: Eigenvalue Estimates for Hill's Equation
%	   with Periodic Coefficients, Ap. Math. Sci. \textbf{53},
%	   2601-2608 (2007)
\bibitem{Ya2} Yajima, K.: Schr\"{o}dinger equations with potentials
	     periodic in time, J. Math. Soc. Japan. \textbf{29}, 729--743
	     (1977).
	
%\bibitem{Ya3} Yajima, K.: Resonances for the AC-Stark effect, Comm. Math. Phys., \textbf{87}, 331--352, (1982). 
	     
\bibitem{Ya} Yajima, K.: Schr\"{o}dinger evolution equations with
	    magnetic fields, J. Anal. Math. \textbf{56}, 29--76 (1991). 

%\bibitem{Yo} Yokoyama, K.: Mourre theory for time-periodic systems,
%	    Nagoya Math. J. \textbf{149}, 193-210 (1998)

\end{thebibliography}
\end{document}